\documentclass[twocolumn,aps,prl]{revtex4-1}
\usepackage{graphicx}
\usepackage{ulem}

\begin{document}

\title{ Fermi-Bose Mixtures Near Broad Interspecies Feshbach Resonances}
\author{Jun Liang Song, Mohammad S. Mashayekhi and Fei Zhou}
\affiliation{Department of Physics and Astronomy, The University of British
Columbia, Vancouver, B. C., Canada V6T1Z1}
\date{{\small June 5, 2010}}

\begin{abstract}
In this Letter we have studied dressed bound states in
Fermi-Bose mixtures near broad interspecies resonance, and implications
on many-body correlations.
We present the evidence for
a first order phase transition between a
mixture of Fermi gas and condensate, and
a fully paired mixture where extended fermionic molecules
occupy a single pairing channel
instead of forming a molecular Fermi surface.
We have further investigated the effect of Fermi surface dynamics,
pair fluctuations and discussed
the validity of our results.
\end{abstract}

\maketitle

Since the observations of molecules of Fermi atoms
near Feshbach resonances, fascinating pairing correlations in cold Fermi gases
have been successfully investigated both experimentally
\cite{Regal03,Strecker03,Jochim03,Zwierlein03,Chin04} and theoretically\cite{Holland01,Ho04,Andreev04}.
Near broad resonances where the atom-molecule coupling is very strong,
pair correlations can also be closely related to the ones in
the BEC-BCS crossover theory pioneered
a while ago\cite{Eagles69,Leggett80,Nozieres85}.
Meanwhile, interspecies Feshbach resonances in Fermi-Bose mixtures
of $^6Li$-$^{23}Na$, $^{40}K$-$^{87}Rb$ and $^6Li$-$^{87}Rb$
have been experimentally observed\cite{Stan04,Inouye04,Ferlaino06,Ospelkaus06,Deh08}.
Recently, weakly bound $^{40}K$-$^{87}Rb$ pairs prepared near Feshbach
resonance were further successfully converted into
cold molecules at JILA\cite{Ni08}, which can potentially lead to exciting
opportunities of studying new quantum states of matter\cite{Buchler07}.
Previous theoretical studies on Fermi-Bose mixtures on the other hand
have been mainly focused on narrow resonances or
when the atom-molecule coupling is very weak
\cite{Powell05,Yabu03,Bortolotti06};
phase boundaries in this limit depend on atom-molecule coupling strengths.

Experimentally, creating and probing correlations
in Fermi-Bose mixtures near narrow resonances are more challenging than
near broad resonances.
For a Feshbach resonance with a width $\Delta B$ and
background scattering length $a_{BG}$, an effective
resonance energy width
can be introduced as $\Gamma_{res}=
2m_R a^2_{BG} (\Delta\mu \Delta B)^2/\hbar^2$\cite{Pethick08,Leggett80};
here $\Delta \mu$ is the difference in magnet moments between the scattering and molecule channel
and $m_R$ is the reduced mass of a pair of Bose and Fermi atoms.
For interspecies resonances so far confirmed
\cite{Stan04,Inouye04,Ferlaino06,Ospelkaus06,Deh08,Simoni03},
$\Gamma_{res}$ can be a few orders of magnitude bigger than the quantum degeneracy energy
at a density of $10^{14} cm^{-3}$\cite{Stan04}.
So some of well studied resonances
are quite broad and
many-body correlations near these interspecies resonances remain to be thoroughly understood.
One of very fundamental questions we hope to answer in this Letter is
{\it what is the nature of quantum matter near broad interspecies resonances.}
Especially, how do dressed two-body bound states evolve when approaching resonances?
and accordingly what kind of many-body correlations are developed in a quantum Fermi-Bose mixture?
And physics near broad resonances can distinctly differ from that near narrow resonances; some basic concepts introduced for narrow resonances
such as molecular Fermi surfaces
might not be directly applicable here.
Motivated by these considerations,
we carry out a study on Fermi-Bose mixtures near broad interspecies Feshbach
resonances which serves as a potential reference for more sophisticated analyses.
Apart from the phase diagrams in terms of
interspecies scattering lengths $a_{bf}$ and Bose($m_B$)-Fermi($m_F$) mass ratio
${m_B}/{m_F}$, we focus on bound state properties of near-resonance
Fermi-Bose mixtures which can be potentially probed in experiments.
Our results are useful for the understanding of correlations in
$^{6}Li$-$^{23}Na$, $^{6}Li$-$^{87}Rb$, and $^{40}K$-$^{87}Rb$ mixtures.
For simplicity, we employ a simplest one-channel Hamiltonian
which captures most important aspects near broad resonances,

\begin{eqnarray}
&H&=\sum_{\bf k} \epsilon_{\bf k}^F f^\dagger_{\bf k}f_{\bf k}+
\sum_{\bf k} \epsilon_{\bf k}^B b^\dagger_{\bf k}b_{\bf k}
\nonumber \\
&+& \frac{V_{bf}}{\Omega}
\sum_{{\bf k},{\bf k}',{\bf Q}} f^\dagger_{
\frac{m_R}{m_B}
{\bf Q}
+{\bf k}}
b^\dagger_{\frac{m_R}{m_F}{\bf Q}
-{\bf k}}
f_{\frac{m_R}{m_B}
{\bf Q}
+{\bf k}'}b_{
\frac{m_R}{m_F}
{\bf Q}
-{\bf k}'}
\end{eqnarray}
where $f^\dagger_{\bf k}$,$b^\dagger_{\bf k}$ ($f_{\bf k}$, $b_{\bf k}$) are creation (annihilation) operators for Fermi and Bose atoms respectively, and
$\epsilon^{F(B)}_{\bf k}={\hbar^2 |{\bf k}|^2\over 2m_{F(B)}}$ are kinetic energies for fermions
(bosons) and $\Omega$ is the volume. $V_{bf}$ is the strength of
interaction that
is related to interspecies scattering lengths $a_{bf}$ via
\begin{equation}
 {1\over V_{bf}}={m_R\over 2\pi a_{bf}\hbar^2}-{1\over \Omega}\sum_{\bf k}
{1\over  \epsilon_{\bf k}^R};
\label{vren}
\end{equation}
here $m_R=m_B m_F/(m_B+m_F)$, $\epsilon_{\bf k}^R=\hbar^2 {\bf k}^2/2 m_R$.
We assume that the background boson-boson interactions are
are repulsive so that the mixture is stable;
to illustrate the idea,
here we only include interspecies scattering.

\begin{figure}[ht]
\includegraphics[width=\columnwidth]{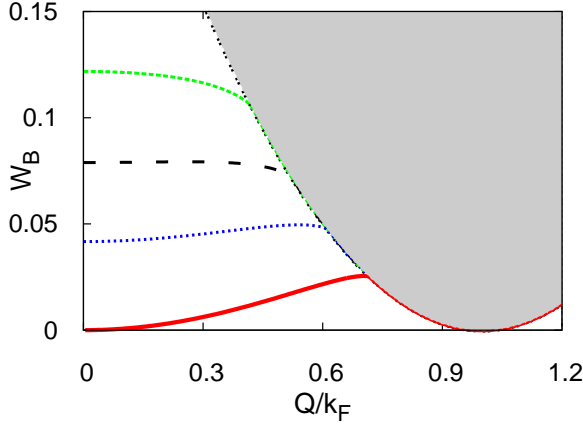}
\caption{(color online) The energy dispersion of bound states
for mass ratio $m_B/m_F=2.175$
or $^{40}K$-$^{87}Rb$ mixtures.
From the top to bottom are $W_B({\bf Q})$ in units of $\epsilon_F^R$
for $1/k_F a_{bf}=-0.2,-0.1145,-0.05,0.0145$.
The shaded region represents pair excitation continuum.
}\label{fig1}
\end{figure}

\begin{figure}[ht]
\includegraphics[width=\columnwidth]{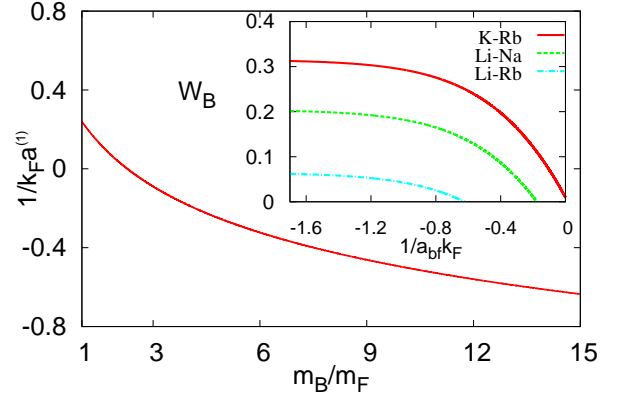}
\caption{(color online) Scattering length $a^{(1)}$
at which the energy cost $W_B$ to create
a molecule with $\hbar{\bf Q}=0$ from a Fermi gas-BEC mixture becomes zero,
as a function of Bose-Fermi mass ratio $m_B/m_F$.
Inset is $W_B$, the energy gap of molecules in units of $\epsilon_F^R$ as a function of $k_F a_{bf}$
(and  $1/a_{bf} < 1/a^{(1)}$)
for different mass ratios or mixtures.}
\label{fig2}
\end{figure}

{\bf Bound states and Pauli blocking effects}
We first consider the binding energy of a pair of Fermi and Bose atoms
with opposite momenta $({\bf k},-{\bf k})$,
in the presence of a condensate(BEC) and
a Fermi surface of Fermi atoms which blocks all states below its Fermi
momentum $\hbar k_F$.
Pauli blocking effects of a Fermi sea indeed
lead to {\it dressed} bound states at arbitrary small negative scattering lengths but with an anomalous dispersion,
or a negative effective mass (see also discussions before Eq.\ref{effM}).
Furthermore, the energy $W_B$ it takes to create a bound state
from a non-interacting ground state
can be either positive or negative depending on scattering lengths $a_{bf}$,
a unique feature of Fermi-Bose systems. This is
because to form a pair of atoms with opposite momenta
$({\bf k},-{\bf k})$ near the Fermi surface $|{\bf k}|=k_F$, a Bose atom has to be promoted to right above
the Fermi surface which results in an energy penalty
$\epsilon^B_F=\hbar^2 k_F^2/2m_B$.
For a bound state with an arbitrary total momentum ${\bf Q}$
or a kinetic energy $\epsilon^C_{\bf Q}=\hbar^2{\bf Q}^2/2(m_F+m_B)$,
the energy cost is $W_B({\bf Q})=\epsilon_F^B+$$\epsilon^C_{\bf Q}+\omega_B$.
The ${\bf Q}$-dependent binding energy
$\omega_B(<0)$ can be obtained by solving the following two-body equation,

\begin{equation}
{-m_R \Omega\over 2\pi a_{bf}\hbar^2} =\Big(
\sum_{|\frac{m_R}{m_B}{\bf Q}+{\bf k}|>k_F}
{1\over \epsilon_{\bf k}^R-
\epsilon_F^R-\omega_B}-\sum_{\bf k} {1\over \epsilon_{\bf k}^R} \Big).
\label{2bd}
\end{equation}
In the limit of small $k_F a_{bf}(<0)$
and when ${\bf Q}=0$, Eq.(\ref{2bd}) leads to
$\omega_B=- 4\epsilon_F^R exp({\pi\over k_F a_{bf}})$,
$\epsilon_F^R=\hbar^2k_F^2/2m_R$.
The dispersion of bound states
that can be probed using photoassociative spectroscopy
is shown in Fig.\ref{fig1}
\cite{Dispersion}.
For small negative scattering lengths $a_{bf}$,
bound states are fully gapped
with positive energies $W_B({\bf Q})$,
and the ground state is a mixture of Fermi gas and BEC.
However, $W_B(0)$, the energy gap of bound states
vanishes at a {\it critical}
scattering length $a^{(1)}$.
In Fig.\ref{fig2}, we present results of $a^{(1)}$ versus $m_B/m_F$.
For heavy Bose atoms, $k_F a^{(1)}$ approaches a small value of
${\pi}/(\ln [m_F/4m_B]+2)$.

To ensure the stability of ${\bf Q}=0$ molecules near the transition line $a^{(1)}$,
we further examine $M_{eff}$, the effective mass near ${\bf Q}=0$.
At scattering lengths $a^{(1)}$ or when $W_B(0)=0$, we find

\begin{eqnarray}
\frac{1}{M_{eff}}=\frac{1}{m_T}[1-\frac{4m_F}{3m_B}g(\frac{m_R}{m_F})];
\label{effM}
\end{eqnarray}
and the dimensionless function $g({x}^2)=x/[(1-x^2)^2(\ln \frac{1+x}{1-x}+\frac{2x}{1-x^2})]$.
As far as $m_B/m_F > 0.7$ and the energy penalty $\epsilon_F^B$ is not too heavy, $M_{eff}$ is
positive, although it can be are much bigger than the bare mass $m_T$($=m_F + m_B$) as a result of
dressing in the Fermi sea.
Below we focus on the limit
of positive $M_{eff}$ that is most relevant to the experimental mass ratio
$m_B/m_F$(between $2.175$ and $14.5$)\cite{Dispersion}.
Note that the binding energy $\omega_B$
is independent of the Bose atom density
when the Fermi sea is treated as a static background.

The above analysis at first sight seems to suggest that
when $W_B$ becomes negative, a small fraction of Fermi and Bose
atoms start forming molecules or a dilute molecular Fermi gas
signifying a phase transition at $a^{(1)}$.
Such a picture was in fact
previously proposed for mixtures near {\it narrow} resonances\cite{Powell05,Yabu03}.
However since the extent of molecules $d_m$
is typically comparable to or much longer than the Fermi wave length $2\pi/k_F$
near broad resonances,
pairs may be accommodated, even before the two-body
gap $W_B$ vanishes,
in other more exotic forms without
forming a molecular Fermi surface.
Below we provide evidence for such a possibility.
Note that a finite two-body gap $W_B$ suggests a local stability
of the Fermi gas-BEC mixture against emergence of
a Fermi gas of molecules, i.e. when $1/a < 1/a^{(1)}$, a molecular Fermi gas can {\it not} be a ground state.
It is in this limit that we illustrate, based on an energetic analysis, that a
{\it third} state or a fully paired state
actually further lowers the energy.

{\bf Energy landscape of pairing states}
Suggested by above discussions,
we consider energetics of pairing states of Fermi-Bose atoms
and from now on focus on {\it homogeneous} mixtures with equal populations of fermions and bosons, i.e. $N_F=N_B$.
Although in principle pairing with a finite total momentum $\hbar {\bf Q}$
can occur in ground states, detailed calculations show that
for a range of mass ratios ($m_B/m_F > 0.2$)
that is relevant to Fermi-Bose mixtures so far studied in experiments,
pairing in $\hbar {\bf Q}=0$ channel
is always dominating and favored, qualitatively consistent with our analyses on the two-body bound states.
Below we only show results of ${\bf Q}=0$ pairing and adopt the simplest pairing wavefunction

\begin{equation}
|g.s.>=exp(c_0b_0^\dagger)\prod_{k\neq0}(u_{\bf k}+v_{\bf k}f^\dagger_{\bf k}b^\dagger_{-\bf k}+\eta_{\bf k}f_{\bf k}^\dagger)|vac>
\end{equation}
where $u_{\bf k}$, $v_{\bf k}$ and $\eta_{\bf k}$ are
three families of variational parameters.
We obtain the energy of the variational states and
 then minimize it with respect to
$u_{\bf k}, v_{\bf k}$ and $\eta_{\bf k}$ that are subject to the normalization condition
$|u_{\bf k}|^2+|v_{\bf k}|^2+|\eta_{\bf k}|^2=1$.
Equilibrium conditions can then be obtained and there are two
solutions for any given ${\bf k}$:
i) a unpaired state with $\eta_{\bf k}=1$ and $u_{\bf k}=v_{\bf k}=0$;
ii) a paired state with $\eta_{\bf k}=0$ and
$v^2_{\bf k}=\frac{1}{2}\left( {1 -\frac{\xi^R_{\bf k}}{E_{\bf k}}}\right)$,
$u^2_{\bf k}=\frac{1}{2}\left( {1 +\frac{\xi^R_{\bf k}}{E_{\bf k}}}\right)$,
$\Delta =-\frac{V_{bf}}{\Omega}\sum_{\bf k} (1-\eta_{\bf k}^2) u_{\bf k} v_{\bf k}$,
where $E_{\bf k}=\sqrt{(\xi^R_{\bf k})^2+4 \Delta^2}$,
$\xi_{\bf k}^{R}=\epsilon_{\bf k}^{R}-\mu$ and
$\mu$ is the pair chemical potential.
Pairing gap $\Delta$, $\mu$ as well as the condensed population $|c_0|^2$
are determined self-consistently,
$N_F =
\sum_{\bf k}\Big(v_{\bf k}^2(1-\eta_{\bf k}^2)+\eta_{\bf k}^2\Big)$,
$N_B =|c_0|^2+\sum_{\bf k}
v^2_{\bf k}(1-\eta_{\bf k}^2)$
and

\begin{figure}[ht]
\includegraphics[width=\columnwidth]{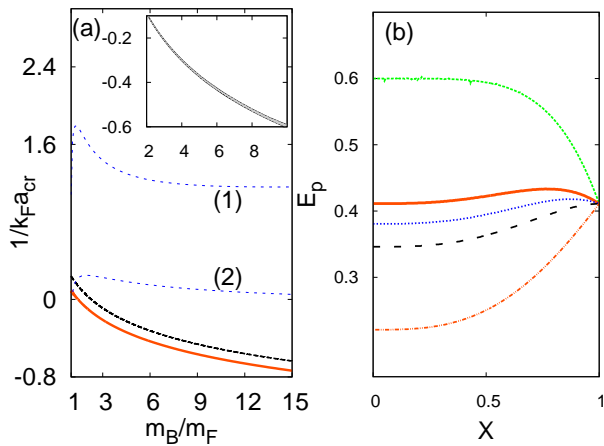}
\caption{(Color online).
(a) $a_{cr}$ at which
a first order phase transition occurs, as a function of mass ratio $m_B/m_F$.
The dashed line is $a^{(1)}$ along which a Fermi gas-BEC
mixture becomes locally unstable.
Near $a_{cr}$, there is a tiny window for phase separation
(shaded area in the inset).
Line 2 (1) is the shifted
$a_{cr}$ ($a^{(1)}$) due to quantum fluctuations
(see discussions around Eq.\ref{GMB}).
(b)$E_p$, energy per pair of Fermi-Bose atoms
in units of $\epsilon_F^R$ as a function of $X$,
the size of Fermi surface of unpaired Fermi atoms
for $m_B/m_F=2.175$ or $^{40}K$-$^{87}Rb$ mixture.
From the top to bottom are $E_p$ for $1/k_F a_{bf}= -5$, -0.1145(for $a_{cr}$), -0.05, 0.0145(for $a^{(1)}$) and 0.2.
$X=1$ corresponds to a unpaired state and
$X=0$ is for a fully paired mixture.
}\label{fig3}
\end{figure}

\begin{eqnarray}
{-m_R\Omega\over 2\pi a_{bf}\hbar^2}
&=& \sum_{\bf k} {1-\eta_{\bf k}^2 \over \sqrt{(\epsilon^R_{\bf k}-\mu)^2+4 \Delta^2}}
-\sum_{\bf k} {1\over \epsilon_{\bf k}^R}.
\label{SCE}
\end{eqnarray}
To minimize the energy, we further choose $\eta_{\bf k}$ to be a step function,
$\eta_{\bf k}=
1$ if $|{\bf k}|\leq X k_F$, and zero otherwise;
the dimensionless variational parameter $X \in [0,1]$
specifies the size of residue Fermi surface of unpaired Fermi atoms.
When $N_F=N_B$, one can also verify that
$X^3$ is equal to $|c_0|^2/N_B$, i.e. the condensation fraction.

\begin{figure}[ht]
\includegraphics[width=\columnwidth]{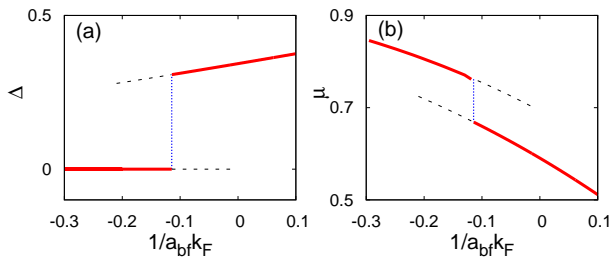}
\caption{(color online).
(a)Pair-breaking energy $\Delta$ and (b)pair chemical potential $\mu$ in units of $\epsilon^R_F$ as a function of scattering length $k_F a_{bf}$
(near $a_{cr}$)
for mass ratio $m_B/m_F=2.175$ or $^{40}K$-$^{87}Rb$.
Dashed lines are for the values of meta-stable states.
The tiny window for phase separation near $a_{cr}$
is too small to show here (see inset of Fig.\ref{fig3}a).
}\label{fig4}
\end{figure}

In Fig.\ref{fig3},\ref{fig4}, we present the main results of variational calculations.
In Fig.\ref{fig3},
the energy per pair of atoms is shown as a function of
$X$, the size of Fermi surface of unpaired fermions.
At small and negative scattering lengths, a Fermi gas-BEC mixture
is a ground state and $X_0=1$, $\mu=\epsilon^F_F+W_B$
$(\epsilon^F_F=\hbar^2 k_F^2/2m_F)$ and $\Delta=0$.
A fully paired state with $X_0=0$ becomes degenerate
with the Fermi gas-BEC mixture ($X_0=1$)
at a critical scattering length $a_{cr}$, beyond which the paired state becomes
a ground state.
However, a Fermi gas-BEC mixture remains to be locally stable until scattering lengths reach the
value of $a^{(1)}$ which is fully consistent with the above study of bound states.
Our variational calculations suggest a
first order phase transition between
a Fermi gas-BEC mixture and
a fully paired mixture.
This later state of
extended molecules is conceptually
different from a Fermi gas of molecules;
instead, all molecules, though fermionic in nature,
occupy the same pairing channel with zero total momentum.
A direct comparison of energies indicates that
a fully paired mixture has lower energies than
a Fermi gas of molecules provided $1/a_{bf} < 1/a^{(2)}$.
For $^{40}K$-$^{87}Rb$ mixtures,
$1/k_F a^{(2)}$ is about $0.25-0.3$
and $1/a^{(2)} > 1/a^{(1)}$;
further towards the molecular side,
a paired mixture is expected to evolve into
a Fermi gas of molecules.

In Fig.\ref{fig4}, we further present the results on the
pair breaking energy $\Delta$ and pair chemical potential $\mu$.
The pair breaking energy can be probed when applying rf pulses to
transfer Fermi atoms to a different hyperfine spin state\cite{Chin04,Baym07}
that weakly interacts with the Fermi-Bose mixture.
The frequency shift in the rf spectroscopy should be
$\hbar \Delta \omega({\bf k})=\frac{1}{2}\big( \xi_{\bf k}^R+\sqrt{|\xi_{\bf k}^R|^2+4\Delta^2} \big).$
In the fully paired phase, Bose atoms are completely depleted and the Bose atom distribution
$n_B({\bf k})$
follows closely the Fermi atom
distribution $n_F({\bf k})$.

{\bf Fermi surface dynamics and pair fluctuations}
At small negative scattering lengths when quantum fluctuations are weak,
our mean-field
analyses on two-body bound states as well as the fully paired states are
asymptotically adequate.
And $a^{(1)},a_{cr}$ estimated in the limit of large mass ratio
($k_F a^{(1)}\sim \pi /\left(\ln [m_F/4m_B]+2\right)$ in Fig.\ref{fig3}) are
quantitatively valid.
However when $k_F a_{\bf}$ is of order of one,
the fluctuations becomes substantial and
the bound states can be further dressed in fluctuating particle-hole pairs;
the energetic analyses are subject to
corrections.
To clarify this, we estimate the dominating effects
of Fermi surface dynamics, i.e. the Gorkov corrections ({\it GMB})
in the two-body scattering vertex due to fluctuating
particle-hole pairs\cite{Gorkov61,Heiselberg00,Pethick08}, and the self-energy ({\it SE}) effect
which mainly represents
the effect of atomic Fermi surface smearing and
the mass renormalization due to scattering by the condensate or Fermi sea.
We then examine the pole structure of the propagator for a pair of Fermi-Bose atoms
taking into account the vertex corrections and self-energy effects
and obtain the binding energy $\omega_B$.
The relative shift caused by the {\it GMB} effect(the leading order term $R_1$)
and the {\it SE} effect(the higher order term $R_2$) is given as
\begin{eqnarray}
\frac{\delta \ln |\omega_B|}{\ln |\omega_B|}=
R_{1}(k_F a_{bf}) +R_2 (k_F a_{bf})^2 \ln |k_F a_{bf}|...
\label{GMB}
\end{eqnarray}
where $R_{1,2}$ both are dimensionless quantities depending on the mass ratio $m_B/m_F$ and can be obtained
numerically\cite{divergency}.
For large mass ratios$(m_B/m_F)$,
one finds that $R_1  \sim -\ln (m_B/m_F)/\pi $ and $R_2 \sim -2/(3\pi^2)$;
and $\omega_B$ is reduced by a factor of $m_F/m_B$
solely due to the {\it GMB} effect.
The net reduction in the binding energy $\omega_B$ leads
to a upward shift of $1/a^{(1)}$ (line 1) in Fig.\ref{fig3}.
In addition, we have estimated that, for the paired state, the amplitude of pair fluctuations $A_1 \sim ({\Delta}/{\epsilon^R_F})^4\sqrt{{\kappa}/{m_R k_F}}$ ($\kappa$ is the compressibility of the paired mixture),
and the corresponding zero point energy (in units of $\epsilon_F^R$) per particle
$A_2 \sim (\Delta/\epsilon^R_F)^4\sqrt{m_R k_F/\kappa}$.
We incorporate these quantum fluctuations (analogous to {\it NSR} effects\cite{Nozieres85})
which further favor pairing, and the {\it GMB} correction into our analysis of energetics of the paired state.
For $^{40}K$-$^{87}Rb$ mixtures, we find that
$A_1 \approx 0.01$, $A_2\approx 0.003$ and
$R_1\approx -0.4$ and $R_2\approx -0.1$; the {\it GMB} and {\it SE}
effects appear to be more dominating.
The modified critical lines
obtained by extrapolating the above analyses to near resonance
are shown in Fig.\ref{fig3} and are
qualitatively consistent with the mean field ones.
Data suggest
that quantum fluctuations tend to enlarge the window between $1/a_{cr}$
and $1/a^{(1)}$ and further stabilize the first order phase transition.
Quantum Monte Carlo simulations similar to those in Ref. \cite{Pilati08} remain to be carried out.

In conclusion, we have examined dressed bound states and provided evidence of a new quantum state of extended molecules near broad interspecies resonances.
As far as three- and higher-body correlations are insignificant, our results can be applied to understand
Fermi-Bose mixtures near broad resonances.
We thank Gordon Baym, Immannuel Bloch, Kirk Madison,
Subir Sachdev, Joseph Thywissen, Jun Ye and Zhenhua Yu for stimulating discussions.
This work is supported by NSERC (Canada) and Canadian Institute for Advanced Research.
{\bf Note added}: Upon the submission of this work, we learned that finite-temperature mixtures near broad resonance were also studied in
Ref.\cite{Fratini10}.


\begin{references}{}
\bibitem{Dispersion}
Far away from resonances or for much smaller Bose-Fermi mass ratios,
the dispersion can be anomalous with a minimum
at $|{\bf Q}|\neq 0$.
\bibitem{divergency}The log-divergency in Eq.\ref{GMB} can be removed if
the compressibility of BEC is taken into account.
\end{references}
\end{document}